\documentstyle[tighten,multicols,eqsecnum,pre,aps]{revtex}

\begin{document}

\draft

\title{Effect of Shear Flow on the Stability of Domains in Two 
Dimensional Phase-Separating Binary Fluids}
\author{Amalie Frischknecht}
\address{Department of Physics, University of California, Santa 
Barbara, California 93106-4030}
\date{\today}

\maketitle

\input{epsf}

\begin{abstract}

We perform a linear stability analysis of extended domains in 
phase-separating fluids of equal viscosity, in two dimensions.  Using 
the coupled Cahn-Hilliard and Stokes equations, we derive analytically 
the stability eigenvalues for long wavelength fluctuations.  In the 
quiescent state we find an unstable varicose mode which corresponds to 
an instability towards coarsening.  This mode is stabilized when an 
external shear flow is imposed on the fluid.  The effect of the shear 
is seen to be qualitatively similar to that found in experiments.

\end{abstract}


\begin{multicols}{2}

\section{Introduction}

Phase separating binary fluids form complex patterns of domains at 
late times after a temperature quench into an unstable state.  The 
morphology of the domains is determined by factors such as the volume 
fractions of the two phases, the viscosities of the two phases, and 
any externally applied forces \cite{siggia,onuki94}.  Of particular 
interest to us is the effect of applying an external shear flow to a 
phase separating binary fluid.  This question is of technological 
importance because many industrial processes involve binary mixtures 
in a flow field.  The final material properties depend on the domain 
morphology, which can be strongly affected by the fluid flow.

At late times after a temperature quench into the two phase region of 
the phase diagram, a phase separating fluid consists of domains of the 
two phases of typical size $R(t)$, which coarsen with time generally 
as a power law $R(t) \propto t^{\alpha}$ \cite{siggia,bray}.  The 
presence of a shear flow dramatically alters the kinetics of the phase 
separation.  The shear flow deforms the domains, interfering with 
their growth so that it competes with the thermodynamic force driving 
the phase separation.  Many theoretical 
\cite{imaeda,onuki86,ohta,padilla} and experimental 
\cite{baum,CPB91,lauger} studies have investigated the effect of the 
shear flow on the growth of the domains and the exponent $\alpha$.  In 
this work we focus on a different aspect of the effect of shear: 
eventually the binary fluid tends towards a dynamic, nonequilibrium 
steady state in which the coarsening instability is stopped by the 
shear flow \cite{onuki86,hashimoto88,Gold:Min93}.  The morphology in 
this stationary state is very anisotropic \cite{baum}.  In relatively 
weak shear, the domains are somewhat deformed, whereas at higher shear 
they can become highly elongated along the flow direction.  A ``string 
phase'' consisting of macroscopically long cylindrical domains forms 
when the two phases are both percolated \cite{string95,string96}.  
This is surprising, since a long cylinder of fluid at rest would 
normally break up via the Rayleigh instability 
\cite{rayleigh,tomotika}, a hydrodynamic instability.  The string 
phase appears to be a fairly robust phenomenon, appearing in both 
critical and off-critical polymer mixtures \cite{string95} and in 
critical micellar solutions \cite{hamano95}.  Thus, the shear flow 
both opposes the thermodynamic instability driving phase separation 
and stabilizes these highly anisotropic domains against hydrodynamic 
instabilities.

Our goal is to understand these stabilizing effects of shear flow.  As 
a first step towards elucidating these effects, we consider a strictly 
two dimensional system.  We expect the operative physical mechanisms 
in the two dimensional fluid to be somewhat different than those in 
the three dimensional case, but the mathematical techniques and 
physical insights developed here will be of use in the future for 
three dimensional calculations.  We consider late times after an 
initial temperature quench into the unstable region of the phase 
diagram, when the system is composed of domains of the two phases 
close to their equilibrium concentrations and separated by 
well-defined interfaces.  We will, however, retain the dynamics of the 
concentration field in our analysis, so that the interfaces between 
domains have a finite width $\xi$.  We model the fluid using the 
coupled Cahn-Hilliard and Navier-Stokes (for creeping flow) equations 
as described in Section \ref{eqtns}.  This is in contrast to the work 
of San Miguel et.  al.  \cite{SanMiguel}, who did an analysis of the 
stability of domains in two dimensional binary fluids, using only the 
Navier-Stokes equation and treating the interfaces as mathematically 
sharp.

\begin{figure}
\begin{center}
\leavevmode
\epsfxsize= 4in
\epsfbox{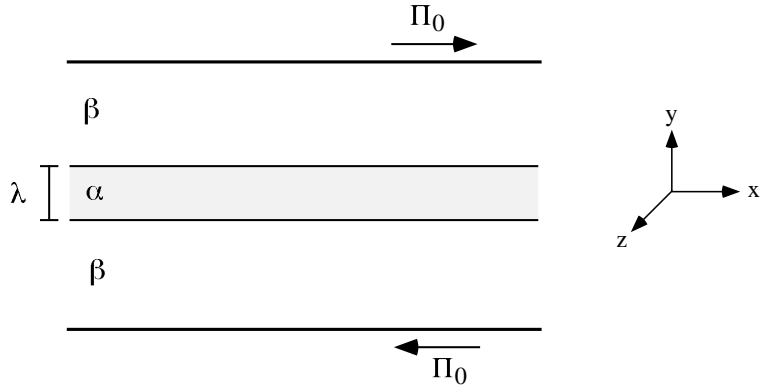}
\end{center}
\caption{\label{lamellafig} Geometry of a single lamellar domain of 
phase $\alpha$}
\end{figure}

In Section \ref{method} we linearize our equations for the general 
case of a system with any number of flat interfaces, and develop some 
useful mathematical machinery.  In Section \ref{1int} we apply our 
methods to the case of a single interface, and reproduce some 
well-known results.  In Section \ref{lamella} we turn to our main 
focus, the stability of a single domain in the form of a strip (in 
three dimensions, a flat sheet) of one phase, immersed in an infinite 
region of the other phase as illustrated in Fig.\ \ref{lamellafig}.  
We impose a shear flow along the $x$-direction by applying a constant 
shear stress $\Pi_{0}$.  In this paper we take the viscosity of the 
two phases to be equal, so that the flow field of the unperturbed 
system is linear.  There are two linearly independent perturbations of 
the lamellar domain along the x-axis.  In the ``zig-zag'' mode the two 
interfaces fluctuate in phase, whereas in the ``varicose'' or 
``peristaltic'' mode they fluctuate out of phase.  We find that in the 
absence of the shear flow the zig-zag mode is stable, whereas the 
varicose mode is unstable to long wavelength perturbations.  We use a 
tight-binding approximation to include the effect of the shear flow.  
In Section \ref{results} we observe that the shear flow mixes the two 
modes so that above a critical shear rate $\dot{\gamma}_{c}$ the 
lamellar domain is stable.  We conclude with some discussion in 
Section \ref{disc}.

\end{multicols}

\begin{multicols}{2}

\section{Model Equations}

\label{eqtns}

We consider a simple binary fluid with one scalar order parameter 
$\Phi$, the difference in concentration between the two components.  
We use the usual Ginzburg-Landau form for the coarse-grained free 
energy of a symmetrical mixture
	\begin{equation}
       F[\Phi]=\int dr \left(\frac{1}{2} K \left(\nabla\Phi\right)^2 - 
       \frac{1}{2}r_o\Phi^2 +  \frac{1}{4}g\Phi^4 \right)\,,
	\end{equation}
where $r_{o}$ and $g$ are positive so that we are below the 
coexistence curve in the two-phase region.  Minimizing the homogeneous 
part of $F$ leads to the values of the concentration in the two bulk 
phases at equilibrium:
\begin{displaymath}
	\Phi = \pm \sqrt \frac{r_{o}}{g} \equiv \pm \phi_{e}\,.
\end{displaymath}
The equation of motion for $\Phi$ is the Cahn-Hilliard equation with 
a convective coupling of $\Phi$ to the velocity field {\bf 
u}:
	\begin{equation}
		\frac{\partial\Phi}{\partial t} = - {\bf u} \cdot \nabla\Phi + 
          M\nabla^2 \frac{\delta F}{\delta\Phi}\,.
         \label{CH}
  \end{equation}
Here $M$ is a concentration-independent mobility.  Since we are 
interested in the late stages of phase separation, we neglect all 
thermal fluctuations.  The equation for the velocity is the 
Navier-Stokes equation for an incompressible fluid, generalized to 
include the coupling of the order parameter to the velocity field  
\cite{chaikin}:
	\begin{equation}
			\rho \frac{\partial {\bf u}}{\partial t} + \rho ({\bf u} 
			\cdot \nabla){\bf u} = \eta \nabla^{2} {\bf u} + \nabla \Phi 
			\frac{\delta F}{\delta\Phi} - \nabla P\,.  
		\label{N-S}
     \end{equation}
In this paper the viscosity $\eta$ will be taken to be independent of 
$\Phi$; hence there is a single viscosity for the fluid independent of 
the concentration pattern.  The pressure $P$ is determined by the 
incompressibility condition
		\begin{equation}
			\nabla \cdot {\bf u} = 0\,.
			\label{incomp1}
     \end{equation}
We will consider only low Reynold's number flow so the convective term 
$({\bf u} \cdot \nabla){\bf u}$ in the Navier-Stokes equation can be 
ignored.  We will also assume that the fluid is sufficiently viscous 
that the velocity responds instantaneously to slow changes in $\Phi$; 
we can then neglect the inertial term $\partial {\bf u}/\partial t$ 
and the resulting equations describe ``creeping'' or Stokes flow.  The 
term coupling the concentration to the velocity field in (\ref{N-S}) 
leads to a capillary force at interfaces, where gradients in $\Phi$ 
induce fluid flow.  Eq.\ (\ref{CH} - \ref{incomp1}) are the same as 
those of ``Model H'' (without the thermal noise terms) used to study 
critical binary fluids \cite{hh}.  These equations have been used 
extensively to study phase separation in binary fluids \cite{bray}.

The first step in a stability analysis is to derive the steady-state 
solutions to the equations of motion.  With the geometry of Fig.\ 
\ref{lamellafig} in mind, we assume that $\Phi$ and {\bf u} are 
functions of $y$ only and look for time-independent solutions.  The 
Cahn-Hilliard equation (\ref{CH}) has steady state solutions 
satisfying
	   \begin{equation}
			\frac{\delta F}{\delta\Phi} = -K \nabla^{2}\Phi - r_{o}\Phi +
			g \Phi^{3} = \mu = {\rm constant}\, ,
			\label{stat sol}
		\end{equation}
where $\mu$ is the exchange chemical potential.  Near a single 
interface, we can take $\mu = 0$ and the concentration has the usual 
``kink'' solution
		\begin{equation}
			\Phi_s = \sqrt{\frac{r_o}{g}} \tanh \sqrt{\frac{r_o}{2K}} y = 
			\phi_{e} \tanh y/\xi \,,
			\label{tanh}
		\end{equation}
where the width of the interface between the two coexisting phases is 
the thermal correlation length $\xi = \sqrt{2K/r_{o}}$.  For a system 
of many lamellar domains, Eq.\ (\ref{tanh}) gives the concentration 
profile at each interface when the domain size is much larger than 
$\xi$.  We note that there is a surface tension associated with the 
presence of an interface, which is just the excess free energy per 
unit area at the interface \cite{langer92}:
\begin{equation}
	\sigma = K \int_{-\infty}^{\infty}dy 
	\left(\frac{d\Phi_{s}}{dy}\right)^{2} = 
	\frac{2\sqrt{2} K^{1/2}r^{3/2}_{o}}{3g} = \frac{2\xi r_{o}^{2}}{3g}\,.
	\label{tension}
\end{equation}
In the stationary state in shear flow there is no velocity in the 
$y$-direction.  We impose a constant shear stress $\Pi_{0}$ so the 
stationary velocity satisfies
      \begin{equation}
      	 {\bf u}_{s} = {\dot \gamma} y {\hat x}
      	\label{}
      \end{equation}
where 
\begin{displaymath}
       \dot{\gamma} \equiv \frac{\Pi_{0}}{\eta} 
\end{displaymath}
is the shear rate.

It is convenient to rewrite our equations in dimensionless form by 
scaling lengths by the correlation length, the concentration by its 
equilibrium magnitude in the bulk phases, and time by the natural 
diffusion time involving the mobility $M$ in the Cahn-Hilliard 
equation.  The velocity is scaled by the correlation length over the 
diffusion time:
  \begin{eqnarray*}
		{\bf \bar{r}} & = & {\bf r} \sqrt{\frac{r_o}{2K}}  =  \frac{{\bf 
		r}}{\xi}\,,  \\
		\bar{t} & = & t \frac{M r_{o}^{2}}{K} = t \frac{2Mr_{o}}{\xi^{2}}, \\       
		\bar{\Phi} & =  & \frac{\Phi}{\Phi_e}\,, \\
		{\bf \bar{u}} & = & {\bf u} \frac{K}{Mr_{o}^{2}\xi} = {\bf u} 
		\frac{\xi}{2Mr_{o}}\,, \\ 
		\bar{P} & = & P \frac{\xi^{2}}{2K\phi_{e}^{2}}\,. 
	\end{eqnarray*}
Note that the new dimensionless correlation length is $\bar{\xi}=1$.  In 
dimensionless form the equations of motion are now
\begin{eqnarray}
	\frac{\partial\bar{\Phi}}{\partial \bar{t}} & = & - 
	{\bf \bar{u}} \cdot \bar{\nabla}\bar{\Phi} + 
	\frac{1}{2}\bar{\nabla}^2\left(-\frac{1}{2}\bar{\nabla}^{2}\bar{\Phi}
	 - \bar{\Phi}+ \bar{\Phi}^{3}\right)\,,  \\
	0 & = & \bar{\nabla}^{2}{\bf \bar{u}} + \frac{1}{\bar{\eta}} 
		\bar{\nabla} \bar{\Phi} \left( 
		-\frac{1}{2}\bar{\nabla}^{2}\bar{\Phi}  - \bar{\Phi}+ \bar{\Phi}^{3}
		 \right) - \frac{1}{\bar{\eta}} \bar{\nabla}\bar{P}\,, \\
    0 & =  & \bar{\nabla} \cdot {\bf \bar{u}}\,.  \label{incomp} 
\end{eqnarray}
We see that the system is characterized by the dimensionless 
parameter, $\bar{\eta}$:
		\begin{equation}
			\bar{\eta} = \frac{Mg\eta}{K} = 
			\frac{4Mr_{o}\eta}{3\sigma \xi}\,. 
		\end{equation}
In dimensionless form, the concentration and velocity profiles derived 
above for a single interface parallel to the flow are
		\begin{equation}
			\bar{\Phi}_{s}(\bar{y})  =  \tanh \bar{y}\,,
		\end{equation}
	\begin{equation}
		\bar{{\bf u}}_s(\bar{y}) = \bar{\dot{\gamma}} \bar{y}\,{\bf \hat{x}}\,.
	\end{equation}
The dimensionless shear rate $\bar{\dot{\gamma}} = \dot{\gamma} 
t_{{\rm diff}}$ is simply the product of the shear rate and the 
diffusion time $t_{\rm {diff}} = \xi^{2}/Mr_{o}$, and thus represents 
a second dimensionless parameter that characterizes the strength of 
the shear flow.

\section{Stability Analysis}

\label{method}

In this section we will develop an overall strategy to examine the 
stability of any number of lamellar domains.  We perform a linear 
stability analysis about the stationary states derived above.  We 
begin by considering small perturbations about the stationary 
solutions (we will drop the bars over the dimensionless variables in 
the rest of the discussion, except where noted):
		\begin{eqnarray}
			\phi & = & \Phi - \phi_s\,,  \\
			{\bf v} & = & {\bf u - u}_s\,.
		\end{eqnarray}
To linear order in the perturbations $\phi$ and {\bf v} the equations 
of motion become
         \begin{equation}
         	\frac{\partial\phi}{\partial t} = -v_y \frac{\partial\phi_s} 
         	{\partial y} - u_s \frac{\partial \phi}{\partial x} + 
         	\frac{1}{2}\nabla^2 
         	\left(-\frac{1}{2}\nabla^2 + W_s(y) \right) \phi\,,  
         	\label{conc}
         \end{equation}
         \begin{eqnarray}	         	
         	0 = \nabla^2{\bf v} + \frac{1}{\bar{\eta}} 
         	\frac{\partial \phi_s}{\partial y} 
         	\left(-\frac{1}{2}\nabla^2+W_s(y) \right) \phi 
         	{\bf \hat{y}} - \nabla P \,,   
         	\label{hydro}
         \end{eqnarray}
         \begin{equation}	        	
         	0 = \nabla \cdot {\bf v} \,.  
         	\label{incomp2}
         \end{equation}
Here $W_s$ is a function of the stationary concentration profile:
	      \begin{equation}
	      	W_s(y) = \left. \frac{\partial^2 f}{\partial \phi^2} 
	      	\right|_{{\displaystyle \phi_s(y)}} = -1 + 3 \phi_{s}^{2}(y)\,.
	      \end{equation}
For a single interface at $y=0$, $W_{s}(y)= 2 - 3{\rm sech}^{2}y$ so that the 
nonconstant part of $W_{s}$ is isolated near the interface.  

In this work we are interested in perturbations along the flow 
direction that are perpendicular to the planar interfaces.  Any such 
perturbation can be written as a sum over Fourier components along the 
x-direction, so we take our perturbations to have the plane-wave forms
\begin{equation}
	\phi = \phi(y){\rm e}^{ikx-\omega t}\,, \quad {\bf v} = {\bf v}(y) 
	{\rm e}^{ikx-\omega t}\,.
	\label{fourier}
\end{equation}
We will consider long wavelength fluctuations for which $k\xi \ll 1$.  
Note that in the following, we take $k$ to be positive, so that $k$ 
represents the magnitude of the wavevector.  First we consider the 
hydrodynamic equations.  If we substitute the expression for ${\bf v}$ 
given in (\ref{fourier}) into the equations of motion for ${\bf v}$, 
Eq.\ (\ref{hydro}) and Eq.\ (\ref{incomp2}), we find we can solve them 
exactly in terms of a Green's function.  First we introduce the stream 
function $\Psi$, defined by
\begin{equation}
	v_x = \frac{\partial \Psi}{\partial y}, \quad v_y = -\frac{\partial 
	\Psi}{\partial x}\,.
\end{equation}
The incompressibility condition (\ref{incomp2}) is then automatically 
satisfied by $\Psi$.  The two components of the Navier-Stokes equation 
(\ref{hydro}) can be used to eliminate the pressure $P$, leaving a 
fourth-order ordinary differential equation for $\Psi = \psi(y) 
\exp(ikx-\omega t)$:
\begin{equation}
	\psi^{(iv)} - 2k^2 \psi^{\prime\prime} + k^4 \psi = 
	\frac{ik}{\bar{\eta}} \phi_{s}^{\prime} \left(\frac{1}{2}k^2\phi - 
	\frac{1}{2}\phi^{\prime\prime} + W_s(y)\phi \right)\,.
	\label{psieq}
\end{equation}
Here primes indicate differentiation with respect to $y$.  The 
boundary conditions are that $\psi$ and its derivative vanish at 
infinity.  This equation can be formally solved using a Green's 
function, to obtain the $y$-component of ${\bf v}$ that is needed in 
the concentration equation (\ref{conc}):
	\begin{eqnarray}
		v_y(y) & = & -ik \psi(y) \nonumber \\ 
		& = & \frac{1}{4\bar{\eta} k} 
		\int_{-\infty}^{\infty} dy' \left(1+k|y-y'|\right) 
		{\rm e}^{\displaystyle{-k|y-y'|}} \phi_s'(y') \nonumber \\ 
		& & \times\left(\frac{1}{2}k^2 \phi(y') - 
		\frac{1}{2}\phi''(y') + W_s(y')\phi(y') \right)\,.
		\label{Greenv}
	\end{eqnarray}
This gives $v_y$ in terms of an integral over $\phi$.

Next substituting (\ref{fourier}) into the concentration equation 
(\ref{conc}) results in an eigenvalue equation for $\omega(k)$:
\begin{eqnarray}
	\omega(k) \phi & = & v_y \frac{d\phi_s}{dy} + ik\dot{\gamma}y 
	\phi \label{eigenval} \\ 
	& &  - \frac{1}{2}\left(\frac{d^{2}}{dy^{2}}-k^{2}\right)
		 \left(-\frac{1}{2} \frac{d^{2}}{dy^{2}}+ \frac{1}{2}k^{2} + 
			W_s(y) \right) \phi \,,  \nonumber
\end{eqnarray}	
where we have used $u_{s}=\dot{\gamma}y$.  A real, positive value of 
$\omega(k)$ indicates stability (damping) of the perturbation.  Note 
that this is essentially an integro-differential equation in which 
$v_{y}$ acts as an integral operator on $\phi$. 

We cannot solve Eq.\ (\ref{eigenval}) exactly so an approximate method 
is needed.  To develop our calculational approach we first consider 
Eq.\ (\ref{eigenval}) without the flow terms:
\begin{equation}
	\omega\phi =  - \frac{1}{2}\left(\frac{d^{2}}{dy^{2}}-k^{2}\right)
	\left(-\frac{1}{2} \frac{d^{2}}{dy^{2}}+ \frac{1}{2}k^{2} + 
	W_s(y) \right) \phi \,.  
	\label{CH2}
\end{equation}
This equation is applicable to the perturbations of domains in a 
binary solid and was used by Langer \cite{langer71} to describe 
coarsening mechanisms in binary alloys. Note that (\ref{CH2}) has the form
	\begin{equation}
			\omega \phi = {\mit \Gamma} F \phi\,, 
		\label{opeq}
	\end{equation}
where we have defined the operators
\begin{mathletters}
\begin{equation}
		{\mit \Gamma} = -\frac{1}{2}\left(\frac{d^{2}}{dy^{2}}- 
		k^{2}\right)
\end{equation}
\begin{equation}
	 F = -\frac{1}{2} \frac{d^{2}}{dy^{2}}+ \frac{1}{2}k^{2} + 
			W_s(y) \,.
\end{equation}
\end{mathletters}
If $\phi_{n}$ is the set of eigenfunctions of (\ref{opeq}) and we 
define a set of ``conjugate'' functions by 
\begin{equation}
	{\mit \Gamma} \tilde{\phi_{n}} = \phi_{n}\,,
\end{equation}
then one can show that ${\mit \Gamma}$ and $F$ are Hermitian operators 
(although their product is not), as long as the $\phi_{n}$ and 
$\tilde{\phi_{n}}$ obey periodic boundary conditions or vanish at 
infinity.  We note that the eigenvalues $\omega_{n}$ are real and the 
eigenfunctions and their conjugates are orthogonal:
\begin{displaymath}
	(\tilde{\phi}_{m}, \phi_{n}) \equiv \int_{-\infty}^{\infty} 
	\tilde{\phi}_{m}^{*}(y) \phi_{n}(y) dy = 0 \quad \rm{for} \quad
	n \neq m \, .
\end{displaymath}
Then for any pair of trial functions $\phi_{0}$ and $\tilde{\phi}_{0}$ 
obeying the same boundary conditions, we can find an upper bound on 
the lowest eigenvalue $\omega$ from a variational relation 
\cite{langer71,jasnow}
	\begin{equation}
		\omega_{min} \leq \frac{(\phi_0, F\phi_0)}{(\tilde{\phi}_0, 
		\phi_0)}\,.
		\label{var}
	\end{equation}
Here the parentheses again indicate inner products.  

To apply Eq.\ (\ref{var}) we need a good trial function $\phi_{0}$.  
It is easy to determine an exact solution of (\ref{CH2}) in the particular 
case when we have a single flat interface present and when $\phi$ is a 
function of $y$ only ($k=0$).  We note that the system is translationally 
invariant, so that any solution that corresponds to a translation of 
the interface by some amount $dy$ is also a solution.  Thus if $y 
\rightarrow y+ dy$ we can write
		\begin{displaymath}
			\phi_s (y+dy) = \phi_s + \frac{d\phi_s}{dy}dy + \cdots
		\end{displaymath}
so it must be that 
	\begin{equation}
	        \phi_0 = \frac{d\phi_s}{dy} = \rm{sech}^{2} \mit{y}
     \end{equation}
is also a solution.  It is easy to verify that this is the case, with 
corresponding eigenvalue $\omega = 0$.  This is the lowest lying 
eigenvalue of Eq.\ (\ref{CH2}) for a system with a single planar 
interface and $k=0$ \cite{jasnow}.  We can use the variational 
principle (\ref{var}) to calculate the stability eigenvalues near this 
$\omega=0$ translational mode for more general situations by assuming 
a trial function formed by appropriate linear combinations of the 
single interface solution \cite{note}.  To use Eq.\ (\ref{var}) we 
also need to determine the conjugate function $\tilde{\phi_{0}}$.  By 
definition the conjugate function satisfies
\begin{equation}
	{\mit \Gamma} \tilde{\phi_{0}} =
	   - \frac{1}{2}\left(\frac{d^{2}}{dy^{2}} - 
	  k^{2}\right)  \tilde{\phi}_{0}(y)  =  \phi_{0}(y)	\,. 
\end{equation}
We can easily solve for $\tilde{\phi}_{0}$ by using a Green's function, 
with boundary conditions that $\tilde{\phi}_{0}$ and 
$\tilde{\phi}^{\prime}_{0}$ vanish at infinity.  We find
\begin{equation}
	\tilde{\phi}_{0}(y) = \int_{-\infty}^{\infty} dy^{\prime} \frac{1}{k}{\rm 
	e}^{\displaystyle{-k\left|y-y^{\scriptstyle \prime}\right|}} 
	\phi_{0}(y^{\prime})\,.
	\label{conj}
\end{equation}
The conjugate function is thus obtained by substituting the desired 
trial function $\phi_{0}$ into Eq.\ (\ref{conj}). 

To summarize the results of this section, we have linearized the 
equations of motion, expressed them parametrically in terms of the 
wavenumber $k$, and solved the hydrodynamic equations for $v_{y}$ as 
an integral over $\phi$.  The eigenvalue equation (\ref{eigenval}) can 
be solved approximately in the absence of the two flow terms (i.e.  
Eq.\ (\ref{CH2})) by evaluating Eq.\ (\ref{var}) using an appropriate 
trial function.  The methods used to include the flow terms will be 
explained in the following sections.

\section{Dispersion Relation for a Single Interface}

\label{1int}

As an example of the variational technique, consider the dispersion 
relation of a single flat interface separating semi-infinite domains 
of the two phases.  We initially neglect hydrodynamic effects and 
focus on solving Eq.\ (\ref{CH2}) for $\omega(k)$.  For a single 
interface located at $y=0$ our trial solution is exactly $\phi_0 = 
\phi_s' = \rm{sech}^{2} \mit{y}$.  There is only one term in 
$F\phi_{0}$, since $\phi_{0}$ is a solution to (\ref{CH2}) for $k=0$:
\begin{eqnarray*}
    F \phi_{0} & = & \left(-\frac{1}{2} \frac{d^{2}}{dy^{2}} + \frac{1}{2} 
       k^{2} + 2 - 3{\rm sech}^{2}y\right) \left({\rm 
       	sech}^{2}y\right) \\
       	& = & \frac{1}{2} k^{2} {\rm sech}^{2}y \,.
\end{eqnarray*}
Using (\ref{conj}) for the conjugate function $\tilde{\phi_{0}}$, we 
find
\begin{eqnarray*}
	\tilde{\phi_{0}}(y) & = & \int_{-\infty}^{\infty} dy' \frac{1}{k} 
	\exp(-k|y-y'|) {\rm sech}^{2}y'  \\ 
	& = & \int_{-\infty}^{\infty} dy' \left( \frac{1}{k} - |y-y'|
	 + \cdots \right)  {\rm sech}^{2}y' \\ 
	& = & \frac{2}{k} - 2\ln \cosh y + O(k) \,,
\end{eqnarray*}
where we have expanded the exponential for small $k$ (long 
wavelengths).  This expansion is not uniform in $y$, but is justified 
since the integrand is only nonzero for small $y'$ and because we will 
only need $\tilde{\phi_{0}}$ for small values of $y$ in the subsequent 
analysis.  The normalization integral is simply
\begin{eqnarray*}
	(\tilde{\phi}_{0}, \phi_{0}) & = & \int_{-\infty}^{\infty} dy \, 
	\left(\frac{2}{k} - 2 \ln\cosh y + O(k)\right) {\rm sech}^{2} y \\
	& = & \frac{4}{k} - 2 (2-2\ln 2) + O(k) \,.
\end{eqnarray*}
Next we apply the variational theorem (\ref{var}) to obtain
    \begin{eqnarray}
       \omega & \leq & \frac{(\phi_{0}, 
       F\phi_{0})}{(\tilde{\phi}_{0},\phi_{0})}
         = \frac{2k^{2}/3}{4/k - 2(2-2\ln 2) + O(k)} \\ \nonumber
        & \cong & \frac{1}{6} k^{3} + O(k^{4}) \,,
        \label{eigen1}
    \end{eqnarray}
where we have retained only the lowest order term in $k$.  If we 
rewrite this relation in dimensional units, we find
        \begin{equation}
        	\omega \cong \frac{1}{3} D k^{3} \xi + O(k^{4})
        \end{equation}
where $D = Mr_{o}$ is a diffusion constant.  This result has been 
obtained previously by Jasnow and Zia \cite{jasnow} and by Shinozaki 
and Oono \cite{oono}.  It also agrees to lowest order in $k$ with the 
perturbative calculation by Bettinson and Rowlands \cite{BR}.  The 
eigenvalue is positive so the single interface is stable, at least to 
long wavelength perturbations.  

The physics here is straightforward.  We know that outside a curved 
interface there is a slight excess concentration, which is given by 
the Gibbs-Thomson relation \cite{langer92}
        \begin{equation}
        	\delta \phi = \frac{\sigma \chi}{R \Delta \phi} \,,
        \end{equation}
where $\sigma$ is the surface tension, $\chi$ is the susceptibility, 
$R$ is the radius of curvature of the interface, and $\Delta \phi = 
2\phi_{e}$ is the miscibility gap.  In our case the curvature of the 
interface is $1/R = Ak^{2}$ where A is the amplitude of the small 
perturbation.  The susceptibility $\chi$ is $\chi^{-1} = 
\partial \mu/\partial \phi = r_{o}$ in the bulk phase.  The 
excess concentration due to the curvature is therefore
\begin{displaymath}
	\delta \phi \sim \frac{Ak^{2}\sigma}{\phi_{e}r_{o}} \sim 
	Ak^{2}\xi\phi_{e} \,,
\end{displaymath}
where we have used (\ref{tension}) to eliminate $\sigma$.  This excess 
concentration will occur outside regions of positive curvature, and 
there will be a corresponding lack of concentration in regions of 
negative curvature, creating a concentration gradient along the 
x-axis.  The flux across the interface caused by this gradient is 
roughly $v \Delta \phi$ where $v$ is the velocity of the interface.  
That velocity, in turn, is just the rate of change of the amplitude 
$A$ of the perturbation, so
 \begin{displaymath}
 	v \Delta \phi = \phi_{e} \frac{dA}{dt} = \phi_{e} \omega A \sim 
 	D \nabla \phi \,.
 \end{displaymath}
The concentration gradient is $\nabla \phi \sim k \delta \phi$; 
putting everything together, we find
 \begin{displaymath}
 	\phi_{e} \omega A \sim  D A k^{3} \xi \phi_{e}
 \end{displaymath}
so that $\omega \sim Dk^{3}\xi$ as advertised.

We can include the lowest order hydrodynamic effects on the dispersion 
relation by performing a perturbative calculation to first order in 
$k$.  We write the full eigenvalue equation (\ref{eigenval}) in the form
	\begin{equation}
			{\mit \Gamma} F \phi + V \phi = \omega \phi \,,
			\label{eigenval2}
		\end{equation}
where the ``unperturbed'' problem is simply Eq. (\ref{CH2}):
\begin{eqnarray*}
	{\mit \Gamma} F \phi_{0} & = & - \frac{1}{2}\left(\frac{d^{2}}
	{dy^{2}}-k^{2}\right)
	\left(-\frac{1}{2} \frac{d^{2}}{dy^{2}}+ \frac{1}{2}k^{2} + 
	W_s(y) \right) \phi_{0} \\ 
	& = & \omega_{0} \phi_{0} \,, 
\end{eqnarray*}
with $\omega_{0} \cong k^{3}/6$ and $\phi_{0} \cong {\rm sech}^{2}y$ from 
the variational result (note these solutions are exact for $k=0$).  
The perturbation $V$ contains the flow terms
\begin{displaymath}
	V = v_y \frac{d\phi_s}{dy} + ik\dot{\gamma} y \phi  \,.
\end{displaymath}
We expect $v_{y}$ to be proportional to a power of $k$ (since for 
$k=0$ there should be no induced velocity in the $y$-direction), so 
$V$ itself is proportional to a power of $k$ and is therefore small 
for long wavelengths.  Expanding $\omega$ and $\phi$ in powers of $k$, 
and multiplying Eq.\ (\ref{eigenval2}) on the left by the 
corresponding left eigenvector $\tilde{\phi}$, one can show in the 
usual way that the lowest order correction to $\omega$ in 
perturbation theory is
	\begin{equation}
		\omega_1 = \frac{\left(\tilde{\phi_0}, V\phi_0 \right)}
		{\left(\tilde{\phi_0}, \phi_0 \right)} \,.
		\label{pert}
	\end{equation}
We solve for the velocity field by substituting $\phi_{0}$ into 
(\ref{Greenv}):
\begin{eqnarray}
	v_{y} & = & \frac{1}{4\bar{\eta} k} 
		\int_{-\infty}^{\infty} dy' \left(1+k|y-y'|\right) 
		{\rm e}^{\displaystyle{-k|y-y'|}} {\rm sech}^{2}y' \nonumber \\
	& &	\times \left(\frac{1}{2}k^2 {\rm sech}^{2}y'  \right) \nonumber \\ 
	 & = & \frac{k}{8\bar{\eta}} \int_{-\infty}^{\infty} dy' 
	 \left({\rm sech}^{4}y' + O(k^{2}) \right) \nonumber \\ 
	 & = & \frac{k}{6\bar{\eta}} + O(k^{2}) \,,
\end{eqnarray}
where we have again expanded the exponential for small $k$.  We find 
that to lowest order $v_{y}$ is linear in $k$, so that overall $V 
\propto k$.  Since our first order perturbative result will only be 
good to $O(k)$, we only need the exact part of the result to the 
unperturbed problem (recall the variational result is $O(k^{3})$), for 
which $\omega_{0}=0$.  In the reference frame in which $u_{s}(y)=\dot{\gamma} 
y$, the integral over the convective term $ik \dot{\gamma} y\phi$ in 
(\ref{pert}) vanishes, so that we obtain a single term in the first 
order correction to $\omega$ from the $v_{y}$ term:
\begin{eqnarray}
	\omega & = & \omega_{0} + \frac{\left(\tilde{\phi_0}, v_{y} 
	\phi_{s}' \right)}{\left(\tilde{\phi_0}, \phi_0 \right)} \nonumber \\ 
	& = &  0+ \frac{\left(\tilde{\phi_0}, \left(k/6\bar{\eta} 
	+O(k^{2})\right) \phi_{0} \right)}
	{\left(\tilde{\phi_0}, \phi_0 \right)} \nonumber \\ 
	& = & \frac{k}{6\bar{\eta}} + O(k^{2}) \,,
\end{eqnarray}
since $\phi_{s}' = {\rm sech}^{2}y = \phi_{0}$ for a single interface. 
If we restore the units in this result we obtain
\begin{equation}
	\omega = \frac{\sigma k}{4\eta} + O(k^{2})
\end{equation}
where $\sigma$ is the surface tension.  This is a well-known result 
for the damping of long wavelength capillary waves on a planar 
interface between two liquids, in the limit that the viscosity is 
sufficiently large that inertial effects can be neglected 
\cite{interface}.

\section{Calculational Method for a Lamellar Domain}

\label{lamella}

We now turn to the stability of a lamellar domain of one phase 
immersed in the other phase, so that we have two interfaces in the 
system as in Fig.\ \ref{lamellafig}.  When the spacing $\lambda$ 
between the two interfaces is at least a few correlation lengths (note 
that we continue to work with scaled variables), $\lambda \gg 1$, the 
stationary concentration profile is
\begin{equation}
	 \phi_{s} = \left\{ \begin{array}{ll} 
	     \tanh (y+\lambda/2) &  -\infty < y < 0  \\    
     	- \tanh (y-\lambda/2) &  0 < y < +\infty  
     	\end{array} \right. 
     	\label{phis}
\end{equation}
where we have arbitrarily taken the $\alpha$ phase with equilibrium 
concentration $\phi_{\alpha} = +1$ to be in the middle, with layer 
thickness $\lambda$.  In this expression we have set the 
exchange chemical potential $\mu$ to zero.  More accurately, we can 
calculate $\mu$ as follows.  The stationary solution that satisfies 
Eq.\ (\ref{stat sol}) is
\begin{displaymath}
	\phi_{s} = \tanh(y+\lambda/2) - \tanh(y-\lambda/2) + \mu \,,
\end{displaymath}
where the regions indicated in (\ref{phis}) are implied.  The chemical 
potential serves as a Lagrange multiplier to keep the concentration 
conserved, so we can find $\mu$ by integrating the concentration field 
over the size of the system and setting it equal to the average 
concentration $\phi_{av}$:
\begin{displaymath}
	\frac{1}{2L} \int_{-L}^{L} \phi_{s}(y) dy = \phi_{av} \,.
\end{displaymath}
We want the volume fraction $x_{\beta}$ of the background phase with 
concentration $\phi_{\beta} = -1$ to be $x_{\beta} = (2L-\lambda)/2L$.  
Using the lever rule
\begin{displaymath}
	x_{\beta} = \frac{\phi_{\alpha}-\phi_{av}}{\phi_{\alpha}-\phi_{\beta}}
\end{displaymath}
and the equilibrium concentrations $\phi_{\alpha} = 1$, 
$\phi_{\beta} = -1$ we find that
\begin{displaymath}
	\phi_{av} = -1 +\frac{\lambda}{L} \,.
\end{displaymath}
Doing the integral over the stationary concentration and keeping only 
the first order corrections in $\exp(-\lambda)$ for $\lambda \gg 1$, 
we find that
\begin{equation}
	\mu \cong -\frac{1}{L} {\rm e}^{-\lambda} \, ,
\end{equation}
so that $\mu \rightarrow 0$ as the system size $L \rightarrow 
\infty$. The dependence of $\mu$ on $\lambda$ will be important to 
our understanding of the physics in Section \ref{noflow} below.

\begin{figure}
\begin{center}
\leavevmode
\epsfxsize=4in
\epsfbox{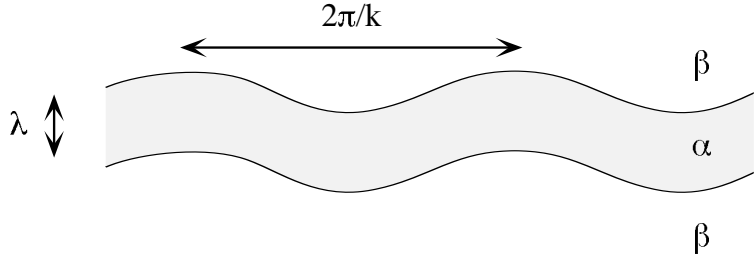}
\end{center}
\caption{\label{zigzagfig} Zig-zag mode}
\end{figure}

\begin{figure}
\begin{center}
\leavevmode
\epsfxsize=4in
\epsfbox{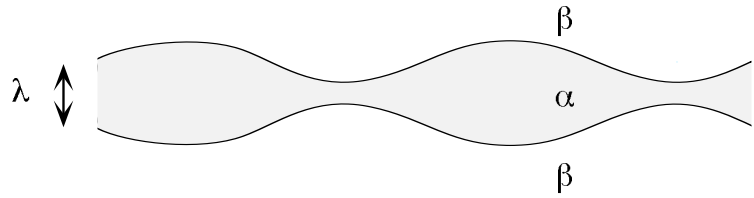}
\end{center}
\caption{\label{varicosefig} Varicose mode}
\end{figure}

Next we want to solve the full eigenvalue equation (\ref{eigenval}) 
for the lamellar domain.  Any perturbation of the domain can be written 
in terms of two linearly independent perturbation modes: either the 
two interfaces can fluctuate in phase with each other to form a 
``zig-zag'' mode, or they can fluctuate out of phase in a ``varicose'' 
or ``peristaltic'' mode.  These modes are pictured in Fig.\ 
\ref{zigzagfig} and Fig.\ \ref{varicosefig}.  Since we are interested 
in calculating the eigenvalues near the marginally stable mode with 
$\omega = 0$ at $k=0$, we take the perturbed concentration field for 
the ziz-zag and varicose modes to be, respectively,
\begin{eqnarray}
   \phi_{z} & = & \frac{1}{2}{\rm sech}^{2}(y+\lambda/2) - 
\frac{1}{2}{\rm sech}^{2}(y-\lambda/2) \,, \label{phizig}  \\
   \phi_{v} & = & \frac{1}{2}{\rm sech}^{2}(y+\lambda/2) + 
\frac{1}{2}{\rm sech}^{2}(y-\lambda/2) \,. \label{varicose}
\end{eqnarray}
The variational theorem (\ref{var}) gives the eigenvalues for these 
two modes in the absence of any hydrodynamic effects.  However, we are 
interested in the effect of the shear flow and of the fluid flow 
induced by gradients in the concentration.  We cannot use the 
perturbation theory approach used in Section \ref{1int} because the 
varicose mode is not a solution to any ``unperturbed'' operator in 
Eq.\ (\ref{eigenval}) (note that the zig-zag mode is the translation 
mode, $\phi_{z}=\phi_{s}'$, and is an exact solution to (\ref{CH2}) at 
$k=0$).  Instead, we adopt a ``tight-binding'' approximation that will 
allow us to solve the full problem.  

\end{multicols}

\begin{multicols}{2}

To implement this approach, we consider the two perturbation modes 
above to be two basis states, and rewrite the eigenvalue equation 
(\ref{eigenval}) as a two-by-two matrix equation in this basis.  We 
use $\phi$ as the right-hand basis state and the conjugate function 
$\tilde{\phi}$ as the left-hand state.  We insert our two trial 
functions (\ref{phizig}) and (\ref{varicose}) for $\phi$ into the 
eigenvalue equation, multiply on the left by the corresponding 
$\tilde{\phi}$, and integrate over all $y$.  In vector notation, we 
have
\begin{eqnarray}
	\phi_{0} & = & \left( \begin{array}{c}
	   a ({\rm sech}^{2}(y+\lambda/2) - {\rm sech}^{2}(y-\lambda/2))/2 \\
	   b ({\rm sech}^{2}(y+\lambda/2) + {\rm sech}^{2}(y-\lambda/2))/2
	   \end{array} \right) \nonumber \\
	   & = &
	   \left( \begin{array}{c}
	   a \phi_{z} \\
	   b \phi_{v}
	   \end{array} \right)
	   \label{basis}
\end{eqnarray}
as our trial function, where $a$ and $b$ are the amplitudes of the 
two modes.  Substituting into Eq.\ (\ref{eigenval}) gives the matrix 
equation
\end{multicols}
\begin{eqnarray}
	\lefteqn{ \left(\begin{array}{cc}
	(\tilde{\phi_{z}}, \phi_{z}) \omega & 0 \\
	0 & (\tilde{\phi_{v}}, \phi_{v})\omega
	\end{array} \right)
	\left(\begin{array}{c} a \\ b \end{array} \right)  =  } \nonumber \\ 
	 & & \left(\begin{array}{cc}
	(\tilde{\phi}_{z},v_{y}^{z}\phi_{s}') + 
	(\tilde{\phi}_{z},ik\dot{\gamma} y\phi_{z})+ 
	(\phi_{z},F\phi_{z}) & 
	(\tilde{\phi}_{z}, v_{y}^{v}\phi_{s}')+
	(\tilde{\phi}_{z},ik\dot{\gamma} y \phi_{v})  \\
	(\tilde{\phi}_{v},v_{y}^{z}\phi_{s}')+
	(\tilde{\phi}_{v}, ik\dot{\gamma} y \phi_{z}) &
	(\tilde{\phi}_{v},v_{y}^{v}\phi_{s}')+(\tilde{\phi}_{v}, ik\dot{\gamma} 
	y\phi_{v})+(\phi_{v}, F\phi_{v})
	\end{array} \right)
		\left(\begin{array}{c} a \\ b \end{array} \right) \,. 
	\label{matrix1}
\end{eqnarray}
\begin{multicols}{2}
Here we have used the definition ${\mit \Gamma} \tilde{\phi} = \phi$.  The 
superscript on $v_{y}$ indicates with which perturbation mode the 
velocity field corresponds, so that $v_{y}^{z}$ is the velocity 
induced by the zig-zag mode and $v_{y}^{v}$ the velocity induced by 
the varicose mode.  On the left-hand side of (\ref{matrix1}) we use 
the orthogonality properties
\begin{displaymath}
	(\tilde{\phi}_{z}, \phi_{v}) = (\tilde{\phi}_{v}, \phi_{z}) = 0 \,.
\end{displaymath}
These also apply to the diffusive terms on the right-hand side; this 
procedure thus ensures that in the absence of any flow effects we 
obtain the same eigenvalues $\omega$ as we would from the variational 
theorem (\ref{var}).  We can now solve (\ref{matrix1}) for the 
stability eigenvalues.  Note that all calculations presented below are 
carried out to the lowest possible order in $k$.

\section{Lamellar Domain Results}

\label{results}

\subsection{Without Shear Flow}
\label{noflow}

We consider first the solution of (\ref{matrix1}) in the absence of 
the external shear flow ($\dot{\gamma}=0$).  The only possible 
off-diagonal terms are the ones involving $v_{y}$.  We begin by 
calculating the necessary integrals that form the matrix elements.
  
Using Eq.\  (\ref{conj}) and expanding for small $k$ as in Section 
\ref{1int}, we find the conjugate function for the zig-zag mode is
\begin{displaymath}
	\tilde{\phi_{z}}(y) = -\ln \cosh (y+\lambda/2)  + \ln \cosh 
	(y-\lambda/2) + k\lambda y + O(k^{2}) \,.
\end{displaymath}
The normalization integral is then
\begin{eqnarray}
	(\tilde{\phi}_{z}, \phi_{z}) & = & \int_{-\infty}^{\infty} dy \left\{
	\left(-\ln \cosh (y+\lambda/2) 
       \right. \right. \nonumber \\ 
	&  &  \; \left. \left. + \ln \cosh (y-\lambda/2) + k\lambda y
	+ O(k^{2})\right)  \right. \nonumber \\
    & & \left. \times \left( {\rm sech}^{2}(y+\lambda/2) + 
	{\rm sech}^{2}(y-\lambda/2) \right) \right/2\} \nonumber \\ 
	& = &  2\lambda - 2 -k\lambda^{2} + O(k^{2}) \,. \label{normz}
\end{eqnarray}
Note that the second term on the right hand side of the above is 
negligible for sufficiently large $\lambda$, but not when $\lambda$ is 
of the order of a few correlation lengths.  Since it is reasonable to 
consider the case of $\lambda$ being a few times $\xi$ (recall 
$\xi=1$), we consider $\exp(-\lambda)$ to be a small parameter in the 
calculation, but not $1/\lambda$, so that we retain terms like the 
additive $2$ in (\ref{normz}).  Next, substituting $\phi_{z}$ into Eq.\ 
(\ref{Greenv}) for the velocity field, we find by expanding for small 
$k$ as before,
\begin{equation}
	v_{y}^{z}(y) = \frac{k}{6\bar{\eta}} - \frac{k^{3}}{48\bar{\eta}} 
	\left(4y^{2}+\lambda^{2} + \frac{\pi^{2}}{3} - 2\right) + O(k^{4}) \,.
\end{equation}
Finally, since $\phi_{z} = \phi_{s}'$, there is only one term in 
$F\phi_{z}$:
\begin{equation}
	F \phi_{z} = \frac{1}{4}k^{2}({\rm sech}^{2}(y+\lambda/2) - 
	{\rm sech}^{2}(y-\lambda/2)) \,.
\end{equation}

Now we turn to the varicose mode.  The conjugate function for this 
mode, expanded for small $k$, is
\begin{eqnarray*}
	\tilde{\phi}_{v}(y) & = & \frac{2}{k} - \ln \cosh(y+\lambda/2) - \ln 
	\cosh(y-\lambda/2)  \\
	& & +k\left(y^{2}+ \frac{\lambda^{2}}{4} + 
	\frac{\pi^{2}}{12}\right) + O(k^{2}) \,.
\end{eqnarray*}
This leads to a normalization integral of
\begin{displaymath}
	(\tilde{\phi}_{v}, \phi_{v}) = \frac{4}{k} -2\lambda -2+4\ln 2 + 
	k\left(\lambda^{2}+\frac{\pi^{2}}{3}\right) + O(k^{2}) \,.
\end{displaymath}
We note that the normalization goes to infinity as $k \rightarrow 0$. 
This is the mathematical manifestation of the fact that the varicose 
mode is not allowed at $k=0$ because it does not conserve mass. For 
any nonzero $k$ however there is no problem. The velocity field for 
the varicose mode is given by 
\begin{equation}
	v_{y}^{v}(y) = \frac{4k}{\bar{\eta}}\left(2\lambda -3\right) {\rm e}^{-2\lambda} 
	y - \frac{k^{3}\lambda}{12\bar{\eta}} y + O(k^{2}{\rm e}^{-2\lambda}, 
	k^{4}) \,.
\end{equation}
In this expression we have not included terms of $O(k^{2}{\rm e}^{-2\lambda})$. 
These terms will be negligible compared to terms of $O(k^{3})$ for 
$k > {\rm e}^{-2\lambda}$ (at such small $k$, of course, the linear 
term in $k$ will dominate over any $k^{2}$ terms). We will see below 
that this condition is met for the $k$ values of most interest. 
Finally, for the varicose mode
\begin{eqnarray}
	F\phi_{v} & = & \frac{1}{4}k^{2}({\rm sech}^{2}(y+\lambda/2) + 
	{\rm sech}^{2}(y-\lambda/2)) \nonumber \\
	& & - 3{\rm sech}^{2}(y+\lambda/2) {\rm 
	sech}^{2}(y-\lambda/2) \,,
\end{eqnarray}
so that $F\phi_{v}$ includes an overlap term between the two 
interfaces.

It is fairly simple to show by straightforward integration that the 
off-diagonal terms in (\ref{matrix1}) vanish (for $\dot{\gamma}=0$):
\begin{displaymath}
	(\tilde{\phi}_{z}, v_{y}^{v}\phi_{s}') = 
	(\tilde{\phi}_{v},v_{y}^{z}\phi_{s}') = 0 \,.
\end{displaymath}
This reduces the matrix equation to
\end{multicols}
\begin{equation}
	\left(\begin{array}{cc}
	(\tilde{\phi_{z}}, \phi_{z}) \omega & 0 \\
	0 & (\tilde{\phi_{v}}, \phi_{v})\omega
	\end{array} \right)
	\left(\begin{array}{c} a \\ b \end{array} \right)
	 = \left(\begin{array}{cc}
	(\tilde{\phi_{z}},v_{y}^{z}\phi_{s}')+(\phi_{z},F\phi_{z}) & 0 \\
	0 & (\tilde{\phi_{v}},v_{y}^{v}\phi_{s}')+(\phi_{v},F\phi_{v})
	\end{array} \right)
		\left(\begin{array}{c} a \\ b \end{array} \right)
	\label{matrix2}
\end{equation}
\begin{multicols}{2}
so that we can solve for each eigenvalue separately:
\begin{equation}
	\omega = \frac{(\tilde{\phi},v_{y}\phi_{s}') + 
	(\phi,F\phi)}{(\tilde{\phi},\phi)}
	\label{sep}
\end{equation}
for each mode.  Using the expressions given above we perform the 
remaining integrals to obtain $\omega$ for each mode.

\begin{figure}[hbt]
\begin{center}
\leavevmode
\epsfxsize=4in
\epsfbox{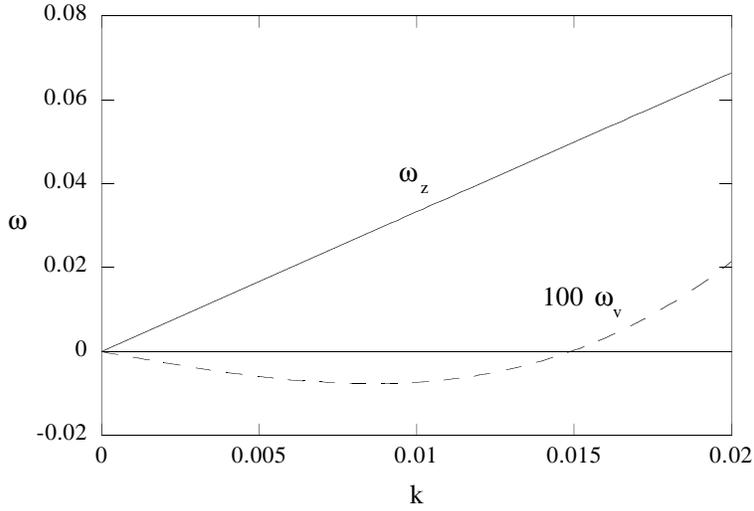}
\end{center}
\caption{\label{omegafig} Dispersion relation for $\dot{\gamma}=0$, with 
$\lambda=6$ and $\bar{\eta} = .1$}
\end{figure}

For the zig-zag mode we find 
\begin{eqnarray}
	\omega_{z} & \cong & \frac{k}{3\bar{\eta}} 
	+ \frac{k^{2}}{6(\lambda-1)} \left(1 
	+ \frac{k\lambda^{2}}{2\lambda -2}\right) \nonumber \\
	& & -\frac{k^{3}}{12\bar{\eta}}\left(\lambda^{2}  
    + \frac{\pi^{2}}{6}-1 + f(\lambda)\right)  + O(k^{4}) \,,
	\label{zig}
\end{eqnarray}
where $f(\lambda)$ is the function
\begin{displaymath}
	f(\lambda) = \frac{\delta_{2}(\lambda)-\lambda\delta_{1}(\lambda) 
	-\delta_{0}}{\lambda -1} \,.
\end{displaymath}
Here $\delta_{0}$ is the definite integral
\begin{displaymath}
	\delta_{0} = \int_{-\infty}^{\infty} dy \, y^{2} {\rm sech}^{2} y 
	\ln \cosh y = 1.70681 \,,
\end{displaymath}
and the functions $\delta_{1}$ and $\delta_{2}$ are the overlap 
integrals
\begin{displaymath}
	\delta_{1}(\lambda) = \int_{-\infty}^{\infty} dy \, y {\rm sech}^{2} y 
	\ln \cosh (y-\lambda) \,,
\end{displaymath}
\begin{displaymath}
	\delta_{2}(\lambda) = \int_{-\infty}^{\infty} dy \, y^{2} {\rm sech}^{2} y 
	\ln \cosh (y-\lambda) \,.
\end{displaymath}
These are integrated numerically using {\em Mathematica}.  We see from 
Eq.\ (\ref{zig}) that the zig-zag mode is stable for small $k$.  The 
terms involving the dimensionless viscosity $\bar{\eta}$ are due to the flow 
field induced by the perturbations in the concentration $\phi$, and 
come from the $v_{y}$ term in (\ref{eigenval}).  Depending on the 
value of $\bar{\eta}$, either the hydrodynamic terms or the diffusive terms 
will dominate.  We find that the stability eigenvalue for the varicose 
mode is
\begin{eqnarray}
	\omega_{v} & \cong & -8k{\rm e}^{-2\lambda} - \frac{12k\lambda 
	{\rm e}^{-2\lambda}}{\bar{\eta}}\left(\frac{2}{3}\lambda -1\right)
	+ \frac{k^{3}}{12} \nonumber \\
	& & + \frac{k^{3}\lambda^{2}}{6\bar{\eta}} + O(k^{2}{\rm 
	e}^{-2\lambda},k^{4}) \,.
	\label{pinch}
\end{eqnarray}
The varicose mode is thus {\em unstable} for sufficiently small 
wavelengths!  The eigenvalues for the two modes are plotted as 
functions of $k$ in Fig.\ \ref{omegafig}.  Here we take $\lambda=6$, 
so that ${\rm e}^{-2\lambda}$ is small (as we assumed above) and 
$k>{\rm e}^{-2\lambda}$ for most of the range in the graph, as 
discussed above. We take the dimensionless viscosity to be $\bar{\eta}=.1$, 
which is a typical value for critical binary fluids.  However the 
overall shape of the dispersion relations remains similar for other 
values of $\lambda$ and $\bar{\eta}$.

\begin{figure}[hbt]
\begin{center}
\leavevmode
\epsfxsize=4in
\epsfbox{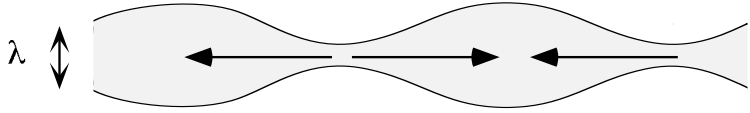}
\end{center}
\caption{\label{pinchfig} Diffusional instability of the varicose mode}
\end{figure}

The instability of the varicose mode may seem unintuitive.  We first 
note that it is unstable only for sufficiently small $k$, and is 
stabilized at larger wavenumbers by the $k^{3}$ curvature term, the 
same term that was obtained in Eq.\ (\ref{eigen1}) for a system with a 
single interface.  Second, the instability is exponentially small in 
the separation between the interfaces, $\lambda$.  This is thus a very 
weak instability.  It is due to a coarsening effect (essentially 
Ostwald ripening) in which thin regions of the middle phase shrink in 
favor of fatter regions.  Recall that the chemical potential $\mu \sim 
{\rm e}^{-\lambda}$.  If $\lambda$ decreases in a region, $\mu$ 
increases, so the chemical potential is higher in the neck regions 
than in the bulges.  This drives a flux from the necks towards the 
bulges (see Fig.\ \ref{pinchfig}).  We can understand the lowest order 
diffusive effect as follows.  First note that the lowest order 
diffusion term in Eq.\ (\ref{pinch}), with units, is
\begin{displaymath}
	\omega \sim -16 D\frac{k}{\xi}{\rm e}^{-2\lambda/\xi}
\end{displaymath}
As before, we can express the velocity of the interface $\omega A$ as
\begin{displaymath}
	\omega A \phi_{e} \sim D \nabla \phi \sim Dk \delta \phi
\end{displaymath}
since the concentration gradient is along the x-direction. The 
excess concentration added  (subtracted) in the bulk regions of the 
necks (bulges) is essentially
\begin{displaymath}
	\delta \phi \sim \frac{A\phi_{e}}{\xi} {\rm sech}^{2}y/\xi \sim 
	\frac{A\phi_{e}}{\xi} {\rm e}^{-2\lambda/\xi}
\end{displaymath}
so that
\begin{displaymath}
	\omega \sim - D \frac{k}{\xi} {\rm e}^{-2\lambda/\xi}
\end{displaymath}
This implies that a large sheet of one phase immersed in the other 
will break up into cylinders via this instability.  Note that this is 
{\em not} the Rayleigh instability of a long fluid cylinder, in which 
the cylinder is unstable towards long wavelength, axisymmetric 
fluctuations.  That is a hydrodynamic instability that occurs for a 
three-dimensional cylindrical interface because the curvature at the 
necks is higher than at the bulges.  In this two-dimensional 
perturbation mode, the curvature at the necks and bulges is of the 
same magnitude (the extra dimension out of the plane of say Fig.\ 
\ref{pinchfig} does not exist), and so there is no curvature-driven 
instability.  The curvature effect is stabilizing, and it is the 
thermodynamic force driving phase separation that causes the 
instability.
\end{multicols}

\begin{multicols}{2}

\subsection{With Shear Flow}

Next we consider what happens when we include the external shear flow.  
Physically, the shear flow tends to mix the two modes since the top 
interface travels in an opposite direction to the bottom interface.  
We might then expect that at some shear rate, the two perturbation 
modes lose their distinguishing features. 

\begin{figure}[hbt]
\begin{center}
\leavevmode
\epsfxsize=4in
\epsfbox{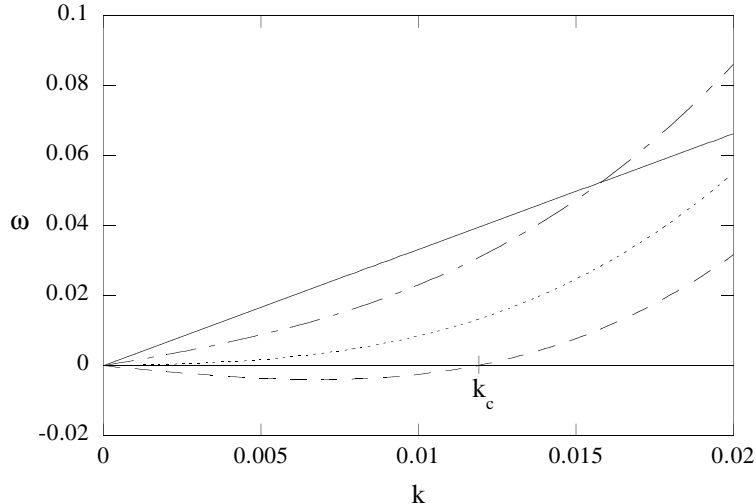}
\end{center}
\caption{\label{omegash} Dispersion relations: $\omega_{+}$ for 
$\dot{\gamma}=.04$ (solid line); $100 \omega_{-}$ for 
$\dot{\gamma}=.04$ (dashed), $\dot{\gamma}=\dot{\gamma}_{c}=.07225$ 
(dotted), and $\dot{\gamma}=.1$ (dash-dot)}
\end{figure}

To calculate the eigenvalues we only need to calculate the matrix 
elements involving the shear.  It is straightforward to show that the 
operator $ik\dot{\gamma}y$ is off-diagonal in the basis of our two perturbation 
modes, i.e.
\begin{displaymath}
	(\tilde{\phi}_{z}, ik\dot{\gamma} y\phi_{z}) = (\tilde{\phi}_{v}, ik\dot{\gamma} 
	y\phi_{v}) = 0 \,.
\end{displaymath}
These two off-diagonal elements are found to be
\begin{eqnarray}
	(\tilde{\phi}_{z},ik\dot{\gamma} y\phi_{v}) & = & ik\dot{\gamma} 
	(\lambda-\lambda^{2}+\delta_{1}) + ik^{2}\dot{\gamma} 
	\lambda\left(\frac{\lambda^{2}}{2} + \frac{\pi^{2}}{6}\right) 
	\nonumber \\
	& & + O(k^{3}) \,,  \\
	(\tilde{\phi}_{v},ik\dot{\gamma} y\phi_{z}) & = & -2i\dot{\gamma} 
	\lambda + ik\dot{\gamma}(\lambda -2\lambda \ln 2 - 
	\delta_{1}+\lambda^{2}) \nonumber \\
	& & - ik^{2}\dot{\gamma} 
	\left(\frac{\lambda^{3}}{2} + \frac{\pi^{2}\lambda}{3}\right) + 
	O(k^{3}) \,.
\end{eqnarray}
\end{multicols}
The stability eigenvalues are now found by diagonalizing Eq.\ 
(\ref{matrix1}), which means solving the secular equation
\begin{equation}
	\left| \begin{array}{cc}
	\omega_{z}-\omega & (\tilde{\phi}_{z},ik\dot{\gamma} 
	y\phi_{v})/(\tilde{\phi}_{z}, \phi_{z}) \\
	(\tilde{\phi}_{v},ik\dot{\gamma} y\phi_{z})/(\tilde{\phi}_{v}, 
	\phi_{v}) & \omega_{v} - \omega  \end{array} \right| = 0 \,.
\end{equation}
Solving for $\omega$ gives
\begin{equation}
	\omega_{\pm}(k) = \frac{1}{2}(\omega_{z}(k) + \omega_{v}(k)) \pm \frac{1}{2} 
	\sqrt{\left(\omega_{z}(k) - \omega_{v}(k)\right)^{2} - \dot{\gamma}^{2} s(k)} 
	\,,
	\label{disp}
\end{equation}
where $s(k)$ is given by
\begin{equation}
	s(k) = 
	\frac{k^{2}\lambda(\lambda^{2}-\lambda-\delta_{1})}{\lambda-1} + 
	\frac{k^{3}}{2(\lambda-1)}\left[ 
	\left(\delta_{1}+\frac{\lambda^{3}}{\lambda-1}\right) 
	(\lambda^{2}-\lambda-\delta_{1}) - \lambda^{4} - 
	\frac{\pi^{3}}{3}\lambda^{2} \right] +O(k^{4}) \,.
\end{equation}
\begin{multicols}{2}
Some examples of the two curves $Re(\omega_{\pm}(k))$ are shown in 
Fig.\ \ref{omegash}.  The spacing between the two interfaces is 
$\lambda = 6\xi$ and we have taken $\bar{\eta}=.1$.  For $\dot{\gamma}=0$ it is clear 
that (\ref{disp}) reduces to our previous results, with 
$\omega_{+}=\omega_{z}$ and $\omega_{-}=\omega_{v}$.  Fig.\ 
\ref{omegash} shows $\omega_{-}$ for three different shear rates (it 
turns out that the curves for $\omega_{+}$ for these same shear rates 
are nearly indistinguishable, so they are plotted as one curve in 
Fig.\ \ref{omegash}).  We see that at low shear rates the unstable 
mode still exists, but the window of wavenumbers over which 
$Re(\omega_{-})<0$ becomes smaller as $\dot{\gamma}$ increases.  Above 
some critical shear rate $\dot{\gamma}_{c}$, the previously unstable mode 
becomes stable for all $k$.  The shear flow thus completely stabilizes 
the varicose mode, by mixing it with the stable zig-zag mode.

\end{multicols}
\begin{multicols}{2}

We can easily solve for the critical shear rate $\dot{\gamma}_{c}$.  First note 
that the first term in (\ref{disp}) is positive, because the negative 
terms in $\omega_{v}$ are exponentially small in $\lambda$.  As $\dot{\gamma}$ 
is increased, the square root term in (\ref{disp}) becomes smaller.  
The effect is that the value of $k$ below which $\omega_{-}<0$ becomes 
smaller with increasing shear; the domain is only unstable to longer 
and longer wavelength perturbations as the shear rate is increased.  
For a given shear rate, $\omega_{-}>0$ for all $k>k_{c}$ where $k_{c}$ 
satisfies $\omega_{-}(k_{c})=0$:
\begin{equation}
	\omega_{z}(k_{c}) + \omega_{v}(k_{c}) = \sqrt{\left(\omega_{z}(k_{c}) - 
	\omega_{v}(k_{c})\right)^{2} - \dot{\gamma}^{2} s(k_{c})} \, .
	\label{crit}
\end{equation} 
The unstable mode becomes stable for all wavenumbers $k$ when 
$k_{c}\rightarrow 0$.  To find the critical shear rate, we first solve 
(\ref{crit}) for $\dot{\gamma}(k_{c})$:
\begin{equation}
	\dot{\gamma}^{2}(k_{c}) = \frac{-4\omega_{z}(k_{c})\omega_{v}(k_{c})}{s(k_{c})} \,.
	\label{gkc}
\end{equation}
Taking the limit $k_{c} \rightarrow 0$ in Eq.\ (\ref{gkc}) gives the 
critical shear rate for complete stabilization:
 \begin{equation}
 	\dot{\gamma}_{c}^{2} = \frac{4(\lambda-1){\rm e}^{-2\lambda}(8\bar{\eta}+ 
 	4\lambda(2\lambda-3))}{3\bar{\eta}^{2}\lambda\left(\lambda^{2}-\lambda+ 
 	\delta_{1}(\lambda)\right)} \,.
 	\label{gammac}
 \end{equation} 
For the specific values $\lambda=6$ and $\bar{\eta}=.1$, one finds 
$\dot{\gamma}_{c}=.07225$ as indicated in Fig.\ \ref{omegash}.

\begin{figure}[hbt]
\begin{center}
\leavevmode
\epsfxsize= 4in
\epsfbox{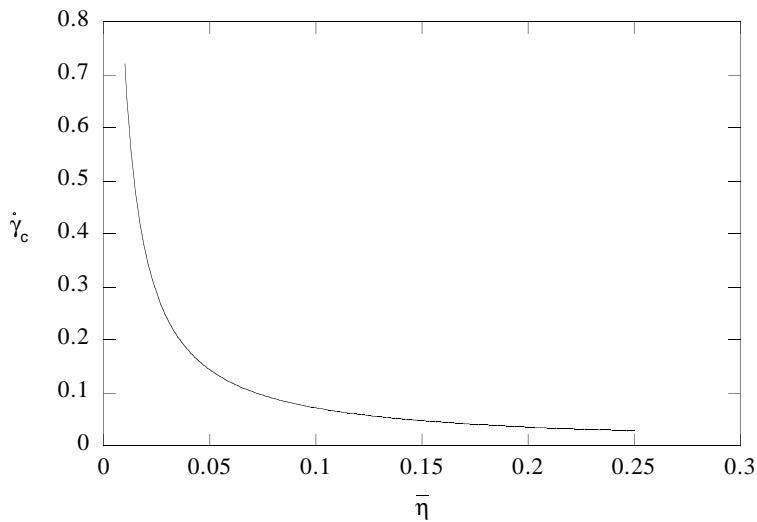}
\end{center}
\caption{\label{gammach} Critical shear rate 
$\dot{\gamma}_{c}(\bar{\eta})$ for $\lambda=6$}
\end{figure}

\begin{figure}[hbt]
\begin{center}
\leavevmode
\epsfxsize= 4in
\epsfbox{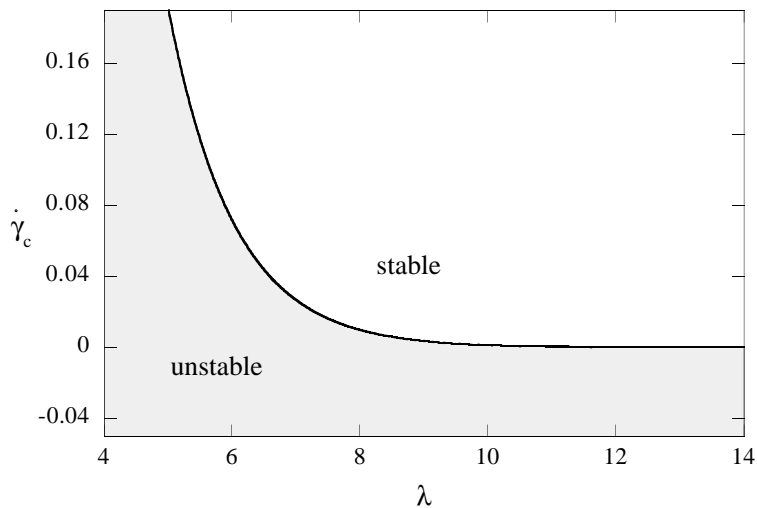}
\end{center}
\caption{\label{gammacl} Critical shear rate 
$\dot{\gamma}_{c}(\lambda)$ for $\bar{\eta}=.1$}
\end{figure}

The critical shear rate is graphed as a function of $\bar{\eta}$ and $\lambda$ 
in Fig.\ \ref{gammach} and Fig.\ \ref{gammacl}.  We note the lamella 
is stable for all $\dot{\gamma} > \dot{\gamma}_{c}$.  We see that $\dot{\gamma}_{c}$ is an 
algebraically decreasing function of $\bar{\eta}$ and an exponentially 
decreasing function of $\lambda$.  Recall that $\bar{\eta} = 4D\eta/3\sigma 
\xi$, so that (\ref{gammac}) tells us that as the viscosity increases, 
or alternatively as the surface tension decreases, the easier it is 
for the shear flow to mix the two modes before the unstable 
perturbation has a chance to grow.  We can also invert (\ref{gammac}) 
to obtain the critical width $\lambda_{c}$ above which the lamella is 
stable for a given shear rate $\dot{\gamma}$.  As we see from Fig.\ 
\ref{gammacl}, given a shear rate $\dot{\gamma}$, at values of $\lambda$ lying 
below the curve the lamella is unstable to the varicose coarsening 
mode whereas for values of $\lambda$ above the curve the lamella is 
stable and will no longer coarsen.  This simple system of a single 
lamellar domain thus exhibits the well-known experimental observation 
that the shear flow tends to halt the phase separation process.

\end{multicols}

\begin{multicols}{2}

\section{Discussion}

\label{disc}

We have seen that in the case of an isolated lamellar domain, shear 
flow has the effect of mixing the zig-zag and varicose modes so that 
they both become stable.  Essentially, the flow eliminates the special 
phase relationship between the two interfaces necessary for the 
varicose mode to exist.  The physics of this mode is that thin regions 
evaporate in favor of thick regions, but in the presence of shear thin 
and thick regions do not exist long enough for this diffusion to take 
place since the fluctuations are being carried downstream.

We would expect that a similar mechanism would apply to a large stack 
(along the $y$-direction) of lamellar domains.  Although the stability 
eigenvalues have not been calculated for this case, the effects seen 
in the single lamellar domain should apply.  Coarsening in the $y$ 
direction in a stack of lamellae is also dependent on thinner regions 
evaporating, their atoms diffusing across the intervening phase to a 
thicker region.  From \cite{langer71} we expect this coarsening 
instability to also have a rate that is exponentially small in 
$\lambda$.  When one considers sinusoidal perturbations of the layers 
in a shear flow, once again the phase relations between interfaces 
will be constantly changing.  As $\lambda$ increases, the atoms must 
diffuse farther across a layer for the pattern to coarsen, but they 
must be able to do so before they are swept downstream by the shear 
flow to a new $x$ position where the diffusion is no longer favored.  
We might anticipate then that in a general two-dimensional system with 
many lamellar domains, for any given shear rate $\dot{\gamma}$ there 
is an upper limit $\lambda_{c}$ to the layer spacing for which the 
coarsening instability is still present.  The shear flow destroys the 
correlations between interfaces necessary for the coarsening 
instability to operate, leading to a dynamic steady state.  The 
strength of the shear flow would determine the typical lamellar width 
$\lambda_{c}(\dot{\gamma})$ present in the system at steady state.

This behavior is qualitatively similar to that seen in the fully three 
dimensional ``string'' phase in shear flow.  We do not expect 
quantitative agreement, however, because the stability analysis of the 
lamellar domain considered here is strongly dependent on the 
dimensionality.  The instability of a long cylinder is much stronger 
than the weak exponential (2D) instability found here.  For the case of a 
viscous cylinder of fluid immersed in another viscous liquid, the 
hydrodynamic instability corresponding to a varicose perturbation 
has a dispersion relation that behaves as \cite{tomotika}
\begin{displaymath}
	\omega \sim -\frac{\sigma}{2\eta a}f(ka)
\end{displaymath}
where $a$ is the radius of the cylinder. Thus, we might expect more 
dramatic effects in this case.

In summary, we have shown that a long extended domain in the two-phase 
state of a two-dimensional, phase-separating binary fluid can be 
stabilized by an applied shear flow.  There is a critical shear rate 
below which the extended domain is unstable towards long wavelength 
fluctuations and above which we predict complete stabilization.  This 
is in qualitative agreement with experiments on dynamic steady states 
in phase-separating fluids under shear flow, however the mechanisms 
operative here are different due to the reduced dimensionality.  We 
intend to report results of a similar calculation for a long 
cylindrical domain under flow in the future.

\section*{Acknowledgements}

I would like to thank J. S. Langer for innumerable helpful discussions 
and support, and G. H. Fredrickson for helpful discussions and a 
thorough reading of the manuscript.  I would also like to thank the 
University of California, Santa Barbara for a Doctoral Scholars 
Fellowship.  This work was supported by the MRL Program of the 
National Science Foundation under Award No.\ DMR 96-32716 and by the 
U.S. DOE Grant No. DE-FG03-84ER45108.

\end{multicols}
\end{document}